\documentclass[aps,10pt,reprint,twocolumn]{revtex4}

\usepackage{epsfig}
\usepackage{graphicx}
\usepackage{dcolumn}
\usepackage{amsmath}
\usepackage{latexsym}
\usepackage{xcolor}
\usepackage{subfig}
\usepackage{mathtools}
\usepackage{footnote}
\usepackage{hyperref}
\hypersetup{
    colorlinks=true,
    linkcolor=blue,
    citecolor=blue,
    filecolor=blue,      
    urlcolor=blue,
    hyperfootnotes=true
}
 
\begin{document}

\title{Vortices in a rotating holographic superfluid with Lifshitz scaling}  

\author{Ankur Srivastav\footnote{
\href{mailto:ankursrivastav@bose.res.in}{ankursrivastav@bose.res.in}}, Sunandan Gangopadhyay\footnote{\href{mailto:sunandan.gangopadhyay@gmail.com}{sunandan.gangopadhyay@gmail.com}\\       \href{mailto:sunandan.gangopadhyay@bose.res.in}{sunandan.gangopadhyay@bose.res.in}}}
\affiliation{Department of Theoretical Sciences, S. N. Bose National Centre for Basic Sciences,\\ Block-JD, Sector-III, Salt Lake City,\\ Kolkata 700106, India}

\begin{abstract}
\noindent We have extended our previous work \cite{anku} on rotating holographic superfluids to include Lifshitz scaling. Presence of this scaling breaks relativistic invariance of the boundary superfluid system and indicates the existence of a Lifshitz fixed point \cite{klm}. We have analytically shown that we still get same vortex solutions as discovered earlier in \cite{anku}. We have recovered previous results for the case of $z=1$, which restores the relativistic invariance in the holographic superfluid system. However, for $z \neq 1$ this study indicate surprising results regarding dissipation in such a holographic superfluid. We found that higher winding number vortices increase with higher values of imaginary chemical potential for values of $z$ in the open interval (1, 2). This result is remarkable because it asserts that dissipation in the rotating holographic superfluid increases in the presence of Lifshitz scaling.
\end{abstract}

\maketitle


\section{Introduction}
\noindent Applied gauge/gravity duality has been subject of interest for the past two decades \cite{nat, bagg, rgc, sah}. It has been tremendously useful in understanding various strongly coupled condensed matter systems where perturbative techniques of standard quantum field theory have almost no access \cite{zaan}. Apart from condensed matter applications, this duality has provided insights in QCD and cosmology as well \cite{nat}. Holographic superconductor \cite{hhh, hhh1} and superfluid models \cite{cph}, which mimic the properties of unconventional superconductors and superfluids, have been explored extensively over the past few years \cite{gr1, gr2, sg1, gg1, js, assg, asdgsg, rbsg, gg2, gg3}. In particular, vortex structure and its dynamics in holographic superfluids and superconductors were studied in various phenomenological settings on the gravity side \cite{prdR, pmchl, mno, prl, cvj1, cvj2, cvj3, xtzh, cyx, jahm}. Existence of vortices is one of the important properties of superfluids under rotation and a variety of vortices have been observed in experiments \cite{ovlet, gev}. Recently, we have also analysed such a rotating holographic superfluid model where we had built novel vortex solutions and showed that dissipation in this model increases with an increase in imaginary chemical potential \cite{anku}. 

\noindent In condensed matter physics, however, there exists a class of systems which do not have relativistic symmetry and thus shows a dynamic scaling ($z \neq 1$) near phase transition \cite{bu}. For such non-relativistic theories, a gravity dual geometry was constructed in \cite{klm} for $z = 2$ and was subsequently generalised for other values of $z$ \cite{bbp, mt}. These gravity geometries are known as Lifshitz geometries which admit following scaling symmetry, 
\begin{eqnarray}
t \rightarrow \lambda^z t, ~~~~~~~~~~ x^i \rightarrow \lambda x^i~. 
\label{Lifshitz_Scaling}
\end{eqnarray}
Gravity dual models constructed out of this geometry are known as Lifshitz holographic models. Lifshitz holographic models of unconventional superconductors have also been analysed in the past \cite{no, lala}. Our interest in this paper is to generalise vortices built in \cite{anku} for Lifshitz holographic model of rotating superfluids. It should be noted from eq.(\ref{Lifshitz_Scaling}) that Lifshitz holographic model reduces to standard $AdS$ holographic model for $z=1$ where relativistic symmetry is restored. In this paper we have considered a disc of radius $R$ at the spacetime boundary and allowed the superfluid to rotate. As mentioned in \cite{anku}, it gives an equivalent description for the static superfluid in a rotating disc. It is known that rotating holographic superfluid admits a vortex state above a critical value of the rotation. It has been numerically shown in \cite{prdR} that above this critical rotation, vortices get excited in the rotating holographic superfluid system. Here we have analytically studied such vortex structure near to this critical rotation in rotating Lifshitz holographic superfluid. In this model also, we have found that chemical potential needs to be purely imaginary for condensate to be real. A holographic model of QCD has been explored with imaginary chemical potential in \cite{kg} in the past. The dissipative nature of imaginary chemical potential in condensed matter system has also been suggested in a previous study \cite{ck}. 

\noindent In this analysis we have found that vortices remain unaffected by Lifshitz scaling $z$. This imply that we again get the same vortex solutions at the boundary disc as the ones obtained in \cite{anku} for any value of $z$. Also linear relation between winding number of vortices and the angular velocity of the rotating superfluid hold irrespective of the value of dynamical exponent $z$. However, Lifshitz scaling does change dissipative nature of vortex state in this model strongly. For $z=1$, results in this model are in agreement with \cite{anku}. That is increase in imaginary chemical potential reduces higher winding number vortices and thereby reduces dissipation in the system. We have obtained a remarkably opposite behaviour in case of $z \neq 1$. It turns out that for such a non-relativistic situation, increase in imaginary chemical potential increases higher winding number vortices and hence dissipation in rotating Lifshitz holographic superfluids also increases. In other words, key finding of this work is that presence of imaginary chemical potential supports dissipation through vortex state for non-relativistic holographic superfluids whereas for relativistic holographic superfluids it opposes such a dissipation. It should be noted that we have considered $z$ to be in the interval [1,2), that is, our analysis does not capture the behaviour of rotating Lifshitz holographic superfluid with the dynamic scaling exponent $z=2$. This is because for $z=2$ there is logarithmic divergence in the gauge fields at the spacetime boundary which needs separate attention. We have left the analysis of this case for future works. 

\noindent This paper has been organised in the following manner. Section (\hyperlink{sec2}{II}) introduces a holographic superfluid model in a ($3+1$)-dimensional Lifshitz spacetime with a static black hole. Near critical angular velocity for a rotating container, we have built vortex solutions in section (\hyperlink{sec3}{III}). In section (\hyperlink{sec4}{IV}), St{\"u}rm-Lioüville eigenvalue approach has been used to analyse this model in bulk direction. This paper ends with section (\hyperlink{sec5}{V}) where we have discussed our final observations in this study and commented on the results. There are also Appendices containing some plots of $\Omega$ vs $\mu$ for different values of $z$ lying between 1 and 2.
\section{Setting Up Holographic Superfluid Model}
\noindent \hypertarget{sec2}{We} consider the following matter action for a holographic superfluid,
\begin{eqnarray}
\mathcal{S} = \dfrac{\textit{l}^2}{16\pi G e^2} \int_\mathcal{M} d^4x \{-\dfrac{1}{4} F^2 -|D\Psi|^2 - m^2\Psi^2\}
\label{Matter_Action}
\end{eqnarray}
where $\textit{l}$ is the radius of curvature of the spacetime geometry, $e$ is charge, $G$ is Newton's constant, $m$ is mass of the scalar field and $F^2 \equiv F_{\mu\nu}F^{\mu\nu}$. Also Faraday tensor and covariant derivative are given by $F_{\mu\nu} = \partial_{[\mu}A_{\nu]}$ and $D_{\mu} = \nabla_{\mu} - ieA_{\mu}$ respectively. In this paper, we shall be working in the probe limit. In this limit, matter sector is assumed to be non back-reacting with the black hole background. This can be achieved mathematically by rescaling scalar and gauge fields with the charge $e$ as $A_\mu \rightarrow \dfrac{A_\mu}{e}$ and $\Psi \rightarrow \dfrac{\Psi}{e}$, and then taking limit $e \rightarrow \infty$. It is equivalent to set $e=1$ in this model.\\
We study this holographic superfluid model in a ($3+1$)-dimensional black hole spacetime with the scaling symmetry given by eq.(\ref{Lifshitz_Scaling}), where $z$ is known as the dynamical exponent. Such a black hole spacetime is realised by the following metric \cite{bu}, 
\begin{eqnarray}
 ds^{2}=-\dfrac{f(u)}{u^{2z}}dt^{2}+\dfrac{du^{2}}{f(u) u^2}+\dfrac{1}{u^2}(dr^2+r^{2}d\theta^{2})~.
\label{Metric}
\end{eqnarray}
The blackening factor is given by $f(u)=(1-u^{z+2})$. We have set $\textit{l}$ and $16 \pi G$ to be unity for convenience and the bulk direction has been scaled in such a way that $u=0$ is the spacetime boundary and $u=1$ represents the event horizon of the black hole. The boundary coordinates $(r, \theta)$ define a $2$-dimensional flat disc. Notice that setting $z=1$ in the above metric restores $AdS_{(3+1)}$ black hole spacetime structure.\\
We now rewrite the metric in Eddington-Finkelstein (EF) coordinates as below,
\begin{eqnarray}
 ds^{2}=-\dfrac{f(u)}{u^{2z}}dt^{2}-\dfrac{2}{u^{z+1}}dtdu+\dfrac{1}{u^2}(dr^2+r^{2}d\theta^{2})~.
\label{EF_Metric}
\end{eqnarray}
We have redefined the EF-time label to $t$ for notational simplicity. Equations of motion for the matter and the gauge fields are given by,
\begin{eqnarray}
    (D^2 - m^2)\Psi = 0 
\label{Matter_field_eq}
\end{eqnarray}
\begin{eqnarray}
   \nabla_\nu F_{\mu}^{~\nu} = j_{\mu} \coloneqq i\{(D_{\mu}\Psi)^{\dagger} \Psi - \Psi (D_{\mu}\Psi)\} ~.
\label{Gauge_field_eq}
\end{eqnarray}
\noindent We assume no explicit time dependency in this model so that all the fields remain stationary. This assumption is justified because we are interested in equilibrium analysis of the rotating superfluid system. In addition to it, we shall be working in the axial gauge, that is, $A_u = 0$. With these conditions, eq.(\ref{Matter_field_eq}) reduces to the following equation,
\begin{eqnarray}
\{\mathcal{D}(u) + \mathcal{D}(r) + \dfrac{1}{r^2}\mathcal{D}(\theta)\}\Psi(u,r,\theta) = 0~.
\label{matter_field_eq_in_axial_gauge}
\end{eqnarray}
The derivative operators are given by,
\begin{eqnarray}
\nonumber
\hspace*{-5mm} \mathcal{D}(u) \equiv u^{z+1}\partial_u\Big(\dfrac{f(u)}{u^2}\partial_u\Big) + i u^{z+1} \partial_u\Big(\dfrac{A_t}{u^2}\Big) + iu^{z-1}A_t \partial_u - \dfrac{m^2}{u^2} \\ 
\nonumber
\mathcal{D}(r) \equiv \dfrac{1}{r}\partial_r(r\partial_r) - \dfrac{i}{r} \partial_r(r A_r) - iA_r \partial_r - A_r^2 ~~~~~~~~~~~\\
\nonumber
\mathcal{D}(\theta)  \equiv \partial_\theta^{~2} - i(\partial_\theta A_\theta + A_\theta \partial_\theta) - A_\theta ^{~2}~.~~~~~~~~~~~~~~~~~~~
\label{Diffrential_Operators}
\end{eqnarray}
We would like to point out here that the information about dynamical exponent $z$ is solely contained in the derivative operator along bulk direction $u$. The other two derivative operators (along boundary coordinates $r$ and $\theta$) remain same as that were in previous study in $AdS$ black hole spacetime model \cite{anku}.
\section{The Holographic Vortex}
\noindent \hypertarget{sec3}{We} are interested in the equilibrium state of this rotating holographic system where vortices exist. As we know that after a critical value of rotation parameter vortices are expected to appear in the holographic superfluid system, we define a deviation parameter, $\epsilon$, from this critical value of rotation, $\Omega_{c}$, by the following relation, 
\begin{eqnarray}
\epsilon \coloneqq \dfrac{\Omega - \Omega_{c}}{\Omega_{c}} 
\label{deviation_parameter}
\end{eqnarray}
where $\Omega$ is considered to be the constant angular velocity of the disc. It has been argued in \cite{prdR} that the superfluid and the boundary disc have a relative velocity and, hence, a static superfluid in a rotating boundary disc could be replaced by a rotating superfluid in a static disc. We have pursued this latter senario. \\
To study this system very near to critical rotation velocity, we series expand all the fields and currents in the following manner \cite{mno},
\begin{eqnarray}
\Psi(u,r,\theta) = \sqrt{\epsilon}\Big(\Psi_1(u,r,\theta) + \epsilon \Psi_2(u,r,\theta)+...\Big)\\ 
A_\mu(u,r,\theta) = \Big(A_\mu^{(0)}(u,r,\theta) + \epsilon A_\mu^{(1)}(u,r,\theta)+...\Big)\\
j_\mu(u,r,\theta) = \epsilon \Big(j_\mu^{(0)}(u,r,\theta) + \epsilon j_\mu^{(1)}(u,r,\theta)+...\Big)~.
\label{Series expansion in epsilon}
\end{eqnarray}
\subsection{Lowest order solutions near spacetime boundary}
\noindent In axial gauge, the lowest order solutions for gauge fields that generate rotation field and the chemical potential are given by following relations,
\begin{eqnarray}
 A_t^{(0)}(u) = \mu(1-u^{2-z}), ~~~~ (z < 2) \\ A_r^{(0)} = 0, ~~~~ A_\theta^{(0)}(r) = \Omega r^2 ~.~~~~~~~
\label{gauge fields in zeroth order}
\end{eqnarray}
 $A_r^{(0)} = 0$ restricts the superfluid flow in radial direction and $A_\theta^{(0)}$ introduces rotation into the superfluid. It should be noted that $z=2$ case is non-trivial due to logarithmic divergence in the $ A_t^{(0)}(u)$ near the spacetime boundary and needs a separate investigation which is extremely difficult to deal with analytically. Hence, we keep ourselves restricted to values of $z$ in the interval [1,2).\\ 
Considering these lowest order solutions for fields near the spacetime boundary, we rewrite eq.(\ref{matter_field_eq_in_axial_gauge}) for lowest order in $\epsilon$, \begin{eqnarray}
\{\mathcal{D}^{(0)}(u) + \mathcal{D}^{(0)}(r) + \dfrac{1}{r^2}\mathcal{D}^{(0)}(\theta)\}\Psi_1(u,r,\theta) = 0
\label{matter_field_eq_lowest_epsilon_order}
\end{eqnarray}
such that the derivative operators become,
\begin{eqnarray}
\nonumber
\mathcal{D}^{(0)}(u) \equiv u^{z+1}\partial_u\Big(\dfrac{f(u)}{u^2}\partial_u\Big) + i u^{z+1} \partial_u\Big(\dfrac{A_t^{(0)}}{u^2}\Big) ~~~~~~~~~~~~~~~\\ 
\nonumber + iu^{z-1}A_t^{(0)} \partial_u - \dfrac{m^2}{u^2}~~~~~~~~~~~~~~~~~~~~~~~~~~~~~~~~~~\\ 
\nonumber
\mathcal{D}^{(0)}(r) \equiv \dfrac{1}{r}\partial_r(r\partial_r)  ~~~~~~~~~~~~~~~~~~~~~~~~~~~~~~~~~~~~~~~~~~~~~~~~~~\\
\nonumber
\mathcal{D}^{(0)}(\theta)  \equiv \partial_\theta^{~2} - i(\partial_\theta A_\theta^{(0)} + A_\theta^{(0)} \partial_\theta) - A_\theta ^{(0)2} ~.~~~~~~~~~~~~~~~~~~
\label{zeroth_order_Diffrential_Operators}
\end{eqnarray}
Using method of variable separation to solve eq.(\ref{matter_field_eq_lowest_epsilon_order})  and writing $\Psi_1(u,r,\theta)$ as a function of $u$ and $(r,\theta)$ 
separately as below,
\begin{eqnarray}
\Psi_1(u,r,\theta) = \Phi(u) \xi(r,\theta)
\label{first_separation}
\end{eqnarray}
eq.(\ref{matter_field_eq_lowest_epsilon_order}) provides the following separated equations,
\begin{eqnarray}
\mathcal{D}^{(0)}(u)\Phi(u) = \lambda \Phi(u)  
\label{matter_field_eq_in Ads direction}
\end{eqnarray}
\begin{eqnarray}
\{\mathcal{D}^{(0)}(r)+\dfrac{1}{r^2} \mathcal{D}^{(0)}(\theta)\}\xi(r,\theta) = - \lambda \xi(r,\theta)
\label{matter_field_eq_in disc}
\end{eqnarray}
where $\lambda$ is some unknown separation constant. 
Eq.(s)(\ref{matter_field_eq_in Ads direction}, \ref{matter_field_eq_in disc}) are eigenvalue equations with eigenvalue $\lambda$. It has been pointed out earlier that the information about dynamical exponent is only in the equation of motion along bulk direction, that is eq.(\ref{matter_field_eq_in Ads direction}). However, the equation on the boundary disc, that is eq.(\ref{matter_field_eq_in disc}), remains same as in \cite{anku} and hence, needs no separate investigation. 
\subsection{Vortex solution}
\noindent In this subsection we shall write the vortex solutions obtained in \cite{anku} and list important properties associated with these solutions. The vortex solutions are given as,
\begin{eqnarray}
\xi(r,\theta) = \eta_{p,n}(r) e^{ip\theta} = a_0 e^{-\Omega r^2/2} F_{p,n}(r) e^{ip\theta}
\label{final_eta_solution}
\end{eqnarray}
where $p \in \mathcal{Z}$ for single valuedness of the solution and $\lambda = 2\Omega (n+1)$ and, $$F_{p,n}(r) = r^p \big(1 + \dfrac{a_2}{a_0} r^2 + \dfrac{a_4}{a_0} r^4 + ... + \dfrac{a_n}{a_0} r^n\big)~.$$
The coefficients $a_i$ can be determined from a recurrence relation given by eq.(26) of \cite{anku}. 
We make following observations regarding these vortices.
\begin{enumerate}
\item These solutions are rotationally symmetric and are subject to Neumann boundary conditions given by,
\begin{eqnarray}
\partial_r\eta_p|_{r=0} = 0 = \partial_r\eta_p|_{r=R} 
\label{boundary_conditions}
\end{eqnarray}
where $R$ is the radius of the disc boundary. 
\item For the case of $n=0$, these boundary conditions imply the quantisation of the angular velocity via $\Omega = \dfrac{p}{R^2}$ for all the values of $p>1$. 
\item For the case of $n=2$, boundary conditions again restrict $p>1$. However, a linear relation between $\Omega$ and $p$ is obtained for large values of $p$. 
\end{enumerate}
Figure(1), taken from \cite{anku}, shows some of these vortex solutions for $n=0$.
\begin{figure}
\includegraphics[width = \linewidth]{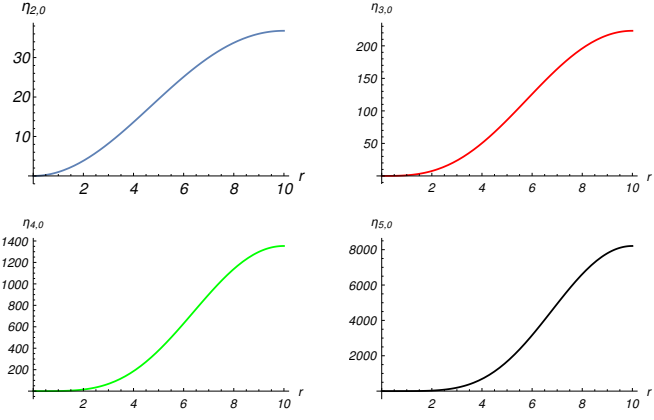}
\caption{Un-normalized lowest order ($n=0$) vortex solutions for different winding numbers. (The value of R is set to be equal to 10).}
\end{figure}
\section{St{\"u}rm-Lioüville Eigenvalue Analysis}
\noindent We shall now solve eq.(\ref{matter_field_eq_in Ads direction}) using St{\"u}rm-Lioüville eigenvalue approach for the eigenvalue $\lambda = 2\Omega$ corresponding to the vortex solutions with $n=0$. Near the critical chemical potential ($\mu \sim \mu_c$), we may take the ansatz  for the lowest order gauge fields, 
\begin{eqnarray}
 A_t^{(0)}(u) = \mu, ~~~ A_r^{(0)} = 0, ~~~ A_\theta^{(0)}(r) = \Omega r^2 ~.
\label{gauge fields approx in zeroth order}
\end{eqnarray}
For simplicity, we shall consider $m^2 = -2z$ and $\Delta = z$. With these considerations, we get,
\begin{eqnarray}
 u^{z+1} \partial_u \Big(\dfrac{1-u^{z+2}}{u^2} \partial_u \Phi(u) \Big) + i u^{z+1} \partial_u \Big(\dfrac{\mu}{u^2} \Phi(u) \Big) \nonumber \\
 + i u^{z-1} \mu \partial_u \Phi(u) + \dfrac{2z}{u^2} \phi = 2\Omega \Phi(u) ~.
\label{bulk equation 01}
\end{eqnarray}
Further we simplify eq.(\ref{bulk equation 01}) as,
\begin{eqnarray}
 u^{z-1}(1-u^{z+2})\partial_u^{2}\Phi-\Big(zu^{2z}+2u^{z-2}-2i\mu u^{z-1}\Big)\partial_u \Phi  \nonumber \\-\Big(2\Omega-\dfrac{2z}{u^2}+2i\mu u^{z-2}\Big) \Phi = 0~.~~ 
\label{bulk equation 02}
\end{eqnarray}
We may now write $\Phi(u)$ near AdS boundary ($u \rightarrow 0$), $$\Phi(u) \simeq~ < \mathcal{O}> u^z \Lambda(u) $$ such that $\Lambda(u)$ is subjected to the following boundary conditions,
\begin{equation}
\Lambda(0) = 1 ~~;~~ \partial_u\Lambda(0)=0~.
\label{bdy cond for Lambda}
\end{equation}
Substituting this form of $\Phi(u)$ in eq.(\ref{bulk equation 02}), we get an equation for $\Lambda(u)$,
\begin{eqnarray}
(1-u^{z+2}) \Lambda^{\prime\prime} +\big(\dfrac{2(z-1)}{u}-3zu^{z+1}+2i\mu\big)\Lambda^{\prime}~~~~~~~~~~~~~~~ \nonumber \\ \hspace*{-5mm}+\big(\dfrac{z(z-3)}{u^2}+\dfrac{2i\mu (z-1)}{u}+\dfrac{2z}{u^{z+1}}-\dfrac{2\Omega}{u^{z-1}}-z^2u^z\big)\Lambda = 0~.~~
\label{Lambda eq 01}
\end{eqnarray}
Here $^{\prime}$ denotes derivative with respect to $u$ in above equation. 
Eq.(\ref{Lambda eq 01}) implies that $\mu$ must be purely imaginary for $\Lambda$ to be real. Hence, we set $Re(\mu) = 0$ and $Im(\mu) = \mu^I$ in above equation. For notational simplicity we shall still denote $\mu^I$ with $\mu$ in the following discussion. With this imaginary chemical potential, eq.(\ref{Lambda eq 01}) takes the following form,
\begin{eqnarray}
(1-u^{z+2}) \Lambda^{\prime\prime} +\big(\dfrac{2(z-1)}{u}-3zu^{z+1}-2\mu\big)\Lambda^{\prime}~~~~~~~~~~~~~~~ \nonumber \\  \hspace*{-5mm}+\big(\dfrac{z(z-3)}{u^2}-\dfrac{2\mu(z-1)}{u}+\dfrac{2z}{u^{z+1}}-\dfrac{2\Omega}{u^{z-1}}-z^2u^z\big)\Lambda = 0~.~~
\label{Lambda eq 02}
\end{eqnarray}
To put eq.(\ref{Lambda eq 02}) in Stürm-Lioüville form, we use the integrating factor,
\begin{eqnarray}
R(u) = u^{z-1} exp(-2\mu \int \dfrac{du}{(1-u^{z+2})})
\label{IF1}
\end{eqnarray}
We may now cast eq.(\ref{Lambda eq 02}) in the Stürm-Lioüville form,
\begin{eqnarray}
(P(u)\Lambda^{\prime}(u))^{\prime}+ Q(u)\Lambda(u) + \Gamma S(u)\Lambda (u) =0~ 
\label{SL equation}
\end{eqnarray}
where eigenvalue $\Gamma = \Omega$ could be obtained using following integral,
\begin{eqnarray}
\Omega = \dfrac{\int_0^1 du (P(u)(\Lambda^{\prime}(u))^2-Q(u)\Lambda^{2} (u))}{\int_0^1 du S(u)\Lambda^{2} (u) }~.
\label{SL eigen01}
\end{eqnarray}
Also, the Stürm-Lioüville coefficient functions $P(u), Q(u)$ and $S(u)$ are given as, 
\begin{eqnarray}
P(u) = u^{z-1}(1-u^{z+2}) R(u) ~\hspace*{42mm}~\nonumber \\  Q(u) = u^{z-1}\big(\dfrac{z(z-3)}{u^2}-\dfrac{2\mu (z-1)}{u}+\dfrac{2z}{u^{z+1}}-z^2u^z\big) R(u)~\nonumber \\
S(u) = -2R(u)~. \hspace*{61mm} 
\label{SL Coefficients}
\end{eqnarray}
Note that the integral in eq.(\ref{IF1}) can be performed exactly to obtain, 
\begin{eqnarray}
R(u) = u^{z-1} exp(-2\mu~ u~ _2F_1(1,\dfrac{1}{z+2};\dfrac{z+3}{z+2};u^{z+2}))~.~
\label{exactIF1}
\end{eqnarray}
$_2F_1(a,b;c;x)$ is the hypergeometric function given by, 
\begin{eqnarray}
 _2F_1(a,b;c;x) =  \sum_{n=0}^{\infty} \dfrac{(a)_n (b)_n}{(c)_n} \dfrac{x^n}{n!}~
 \label{hypergeo}
\end{eqnarray}
where $(m)_n \equiv m(m+1)...(m+n-1)$.\\
We now consider a trial function for $\Lambda(u)$ of the form, $$\Lambda_{\alpha}(u) = (1-\alpha u^2)$$ which satisfies the given boundary conditions, that is, $\Lambda(0) = 1,~ \partial_u\Lambda(0)=0$.
With this trial function, we have to extremize $\Omega_{\alpha}$ with respect to $\alpha$,
\begin{eqnarray}
\Omega_{\alpha} = \dfrac{\int_0^1 du (P(u)(\Lambda_{\alpha}^{\prime}(u))^2-Q(u)\Lambda_{\alpha}^{2} (u))}{\int_0^1 du S(u)\Lambda_{\alpha}^{2} (u) }~.
\label{SL eigen02}
\end{eqnarray}
\subsection{Analysis for z=1}
\noindent Let us first consider the case for dynamical exponent $z=1$. In this case, eq.(\ref{Metric}) shows that bulk spacetime becomes $AdS_{(3+1)}$ black hole spacetime, which is exactly the one that we have analysed in \cite{anku}. Near $AdS$ boundary $(u \rightarrow 0)$ we know that $\Phi(u) \simeq < \mathcal{O}> u \Lambda(u) $, which is the same as in this case with $z=1$. So we get the following Stürm-Lioüville form to solve for, 
\begin{eqnarray}
(P(u)\Lambda^{\prime}(u))^{\prime}+ Q(u)\Lambda(u) + \Gamma S(u)\Lambda (u) =0~ 
\label{SL equation1}
\end{eqnarray}
where eigenvalue $\Gamma = \Omega$ and, 
\begin{eqnarray}
P(u) = (1-u^3) R(u) \nonumber \\  Q(u) = -uR(u)~~~~~~\nonumber \\
S(u) = -2R(u)~.~~~~
\label{SL Coefficients1}
\end{eqnarray}
The integrating factor in this case is given as, 
\begin{eqnarray}
R(u) = exp(-2\mu ~u~_2F_1(1,\dfrac{1}{3};\dfrac{4}{3};u^{3}))~.
\label{IFz1}
\end{eqnarray}
We solve this problem by considering the trial function as discussed above. Also, near $AdS$ boundary $u \rightarrow 0$, we further approximate integrating factor $R(u)$ in the following manner, 
\begin{eqnarray}
R(u) \simeq (1-2\mu ~u~_2F_1(1,\dfrac{1}{3};\dfrac{4}{3};u^{3}))~.
\label{IF3}
\end{eqnarray}
\begin{figure}
\includegraphics[width = \linewidth]{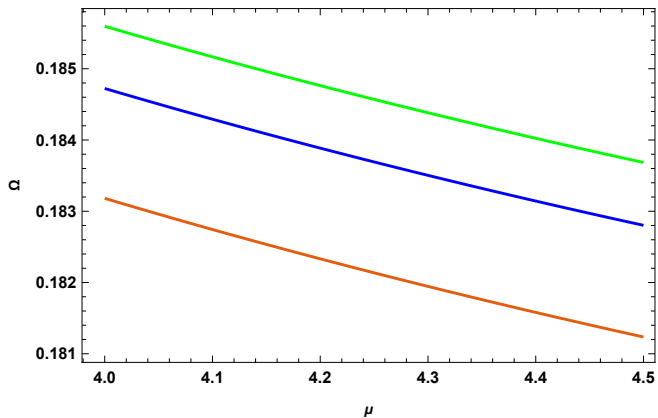}
\caption{$\Omega$ vs $\mu$ for z=1}
\end{figure}
\noindent Figure(2) shows the variation of extremised $\Omega \equiv \Omega_{\alpha=\alpha^*}$ with the increasing value of imaginary chemical potential, $\mu$. Note that three colour plots in the figure represent three different orders upto which we have approximated the hypergeometric function in eq.(\ref{IF3}) for calculations. \textcolor{orange}{Orange} plot is obtained with lowest order approximation while \textcolor{blue}{blue}  and  \textcolor{green}{green} plots result from next consecutive orders of approximation. We shall follow this colour scheme throughout this paper. A few observations are in order regarding this graph which we enumerate below.
\begin{enumerate}
\item This graph between $\Omega$ and $\mu$ shows the same decreasing pattern as in \cite{anku}.   
\item Viewed in conjunction with $\Omega = \dfrac{p}{R^2}$, this analysis shows that for $AdS_{(3+1)}$ holographic superfluid model analysed near equilibrium, presence of $\mu$ opposes the formation of higher winding number vortices.  
\item As is well known in gauge/gravity duality that vortices in holographic superfluid provide mechanism for external perturbations to decay through black hole horizon and hence represent dissipation in such gravity dual systems, this graph suggests that for $z=1$ case, as $\Omega$ decreases with increase in $\mu$, hence $\mu$ supports less dissipation in the system.
\end{enumerate}

\subsection{Analysis for z $\neq$1}
\noindent In this case, the Stürm-Lioüville form of the equation is given by eq.(\ref{SL equation}) and the value of $z$ lies in the interval $(1,2)$. Notice that $z \neq 2$ because of logarithmic divergence of the fields at the boundary $u \rightarrow 0$. With the assumed trial function $\Lambda_{\alpha}(u) = (1-\alpha u^2)$, we need to extremise the following eigenvalue integral, 
\begin{eqnarray}
\Omega_{\alpha} = \dfrac{\int_0^1 du (P(u)(\Lambda_{\alpha}^{\prime}(u))^2-Q(u)\Lambda_{\alpha}^{2} (u))}{\int_0^1 du S(u)\Lambda_{\alpha}^{2} (u) }~.
\label{SL eigen03}
\end{eqnarray}
where 
\begin{eqnarray}
P(u) = u^{z-1}(1-u^{z+2}) R(u) ~~~~~~~~~~~~~~~~~~~~~~~~~~~~~~~~~~~~~\nonumber \\  Q(u) = u^{z-1}\big(\dfrac{z(z-3)}{u^2}-\dfrac{2\mu (z-1)}{u}+\dfrac{2z}{u^{z+1}}-z^2u^z\big) R(u)\nonumber \\
S(u) = -2R(u)~.~~~~~~~~~~~~~~~~~~~~~~~~~~~~~~~~~~~~~~~~~~~~~~~~~~~
\label{SL Coefficients3}
\end{eqnarray}
Here we shall use the following approximated form of eq.(\ref{exactIF1}),
\begin{eqnarray}
R(u) = u^{z-1}(1-2\mu ~ u~ _2F_1(1,\dfrac{1}{z+2};\dfrac{z+3}{z+2};u^{z+2}))~.~~
\label{IF4}
\end{eqnarray}
In Figure(3), we have shown the variation of extremised values of $\Omega$ against imaginary chemical potential $\mu$ for $z= \dfrac{3}{2}$ by considering three orders of approximation of $_2F_1(1,\dfrac{1}{z+2};\dfrac{z+3}{z+2};u^{z+2})$. As we have mentioned, colour codes remain same.
\begin{figure}
\includegraphics[width = \linewidth]{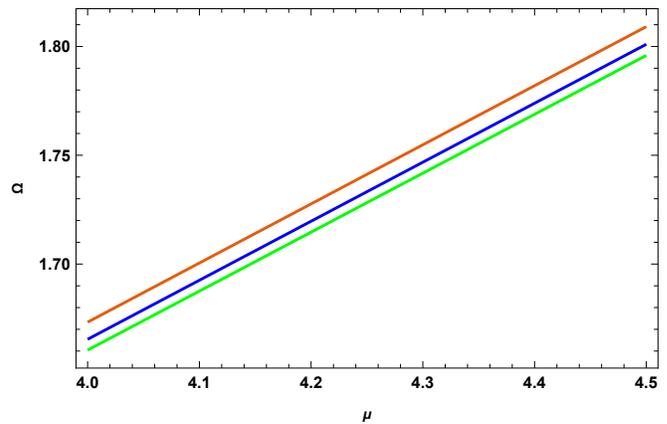}
\caption{$\Omega$ vs $\mu$ for $z=\dfrac{3}{2}$}
\end{figure}
Let us summarise the key observations from this graph.
\begin{enumerate}
\item For $z=\dfrac{3}{2}$, $\Omega$ shows an increasing pattern with $\mu$ unlike in the previous case for $z=1$.   
\item This behaviour implies that for a holographic superfluid with Lifshitz geometry and dynamical exponent $z=\dfrac{3}{2}$, higher winding number solutions are more favourable with increasing value of $\mu$.   
\item In terms of dissipation in such a rotating holographic superfluid, we conclude from this result that higher values of $\mu$ introduce more dissipation in the presence of Lifshitz fixed points (in gauge/gravity duality Lifshitz geometry of the bulk theory is dual to a boundary theory with Lifshitz fixed point.).
\item Same increasing trend for $\Omega$ with $\mu$ is obtained for other values of $z$ in the interval $(1,2)$. Cases with $z= \{1.1, \dfrac{5}{4}, \dfrac{7}{4} \} $ are given in \textit{Appendix I}. 
\end{enumerate}
\section{Conclusion and Remarks}
\noindent In this work we have studied the properties of Lifshitz scaling in the holographic superfluid model under rotation. We have explicitly shown that for $z=1$ our results match with \cite{anku}. Although vortex structure at the boundary disc remains same for all the values of $z$, Lifshitz holographic system differs significantly from the holographic superfluid model in AdS black hole spacetime. In fact, for $1< z < 2$ we get remarkably different trend between $\Omega$ and $\mu$. Our analysis shows that presence of Lifshitz scaling in holographic superfluids does allow vortex formation if we put it under rotation. However, high values of the chemical potential $\mu$ support the formation of higher winding number vortices. This implies that for holographic superfluids with Lifshitz scaling, $\mu$ increases dissipation in the system unlike in the case for AdS black hole model where it has been shown in \cite{anku} that $\mu$ suppresses dissipation by disfavouring the formation of high winding number vortices. Because presence of Lifshitz scaling breaks the relativistic invariance, gravity model built using Lifshitz geometry is dual to non-relativistic boundary superfluid system. In this context we may conclude from this analysis that $\mu$ favours the dissipative vortex state for the non-relativistic superfluids having Lifshitz scaling symmetry whereas for relativistic boundary superfluid systems, presence of imaginary chemical potential opposes the dissipation as $\Omega$ decreases with increase in $\mu$ \cite{anku}. We have also checked the robustness of our results for a different choice of trial function as well. Results obtained in that analysis are given in \textit{Appendix II} which show that the qualitative difference between relativistic and non-relativistic cases remain same.  

\vspace*{2mm}

\noindent {\bf{Acknowledgements}}:  AS would like to acknowledge department of science and technology, government of India for the research fellowship. Authors would also like to thank anonymous referee for sharing some valuable remarks.


\section*{Appendix I: Some more plots of $\Omega$ vs $\mu$ for values of z lying between 1 and 2}
\begin{figure}[h]
\includegraphics[scale=0.75]{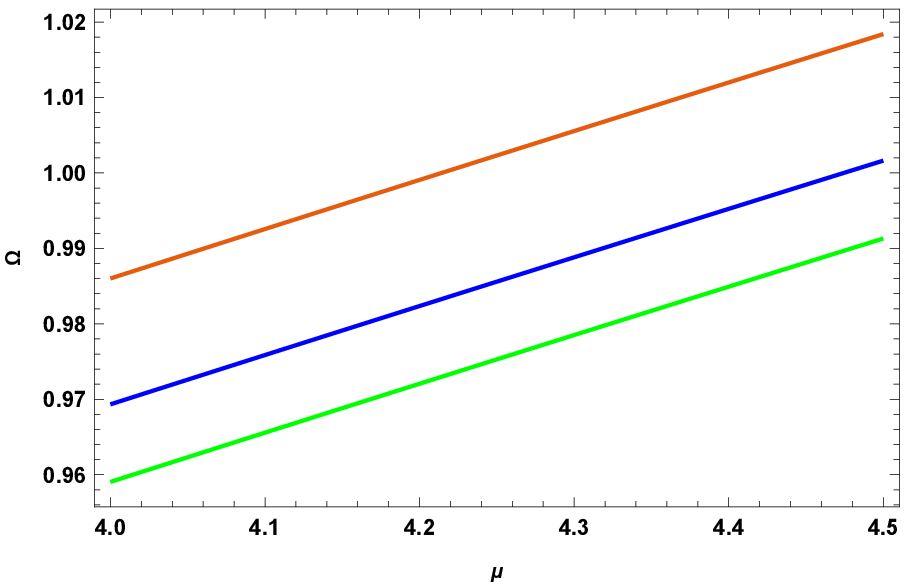}
\caption{$\Omega$ vs $\mu$ for $z=1.1$}
\end{figure}

\begin{figure}[h]
\includegraphics[scale=0.75]{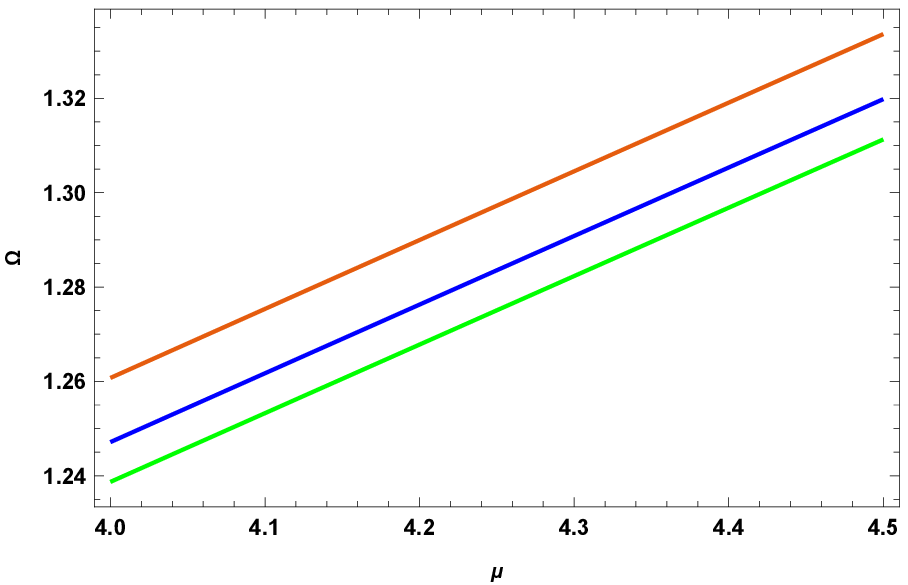}
\caption{$\Omega$ vs $\mu$ for $z=\dfrac{5}{4}$}
\end{figure}

\begin{figure}[h]
\includegraphics[scale=0.75]{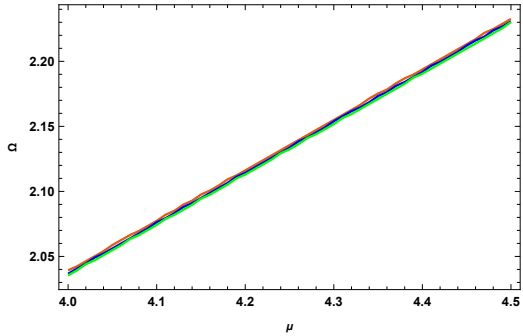}
\caption{$\Omega$ vs $\mu$ for $z=\dfrac{7}{4}$}
\end{figure}

\section*{Appendix II: $\Omega$ vs $\mu$ plots for trial function $\Lambda_{\alpha}(u) = (1-\alpha u^{z+1})$}
\noindent We have considered another trial function, which is also well behaved with the given boundary conditions for $\Lambda_{\alpha}(u)$ given in eq.(\ref{bdy cond for Lambda}). In this case also, we have followed the same analysis as given in Section(IV) and obtained the plots between $\Omega$ and $\mu$ for different values of $z$. We have found that the qualitative behaviour of these plots does not change implying that even with this trial function we get drastically different behaviour for the relativistic ($z=1$) and non-relativistic ($z \neq 1$) holographic superfluids. Below we have provided plots for $z = \{ 1, 1.1,  3/2\}$ for this choice of trial function where colour codes have same meaning as before and show different orders of approximation for hypergeometric function in eq.(\ref{IF4}).

\begin{figure}[h]
\includegraphics[scale=0.75]{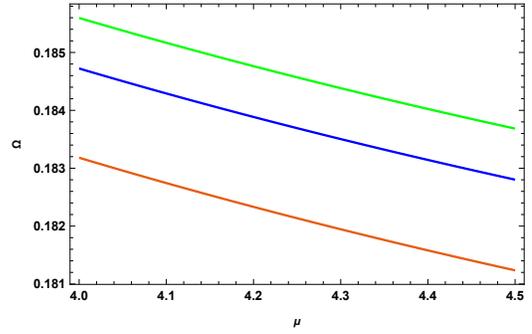}
\caption{$\Omega$ vs $\mu$ for $z=1$}
\end{figure}
\vspace*{-9mm}

\begin{figure}[h]
\includegraphics[scale=0.75]{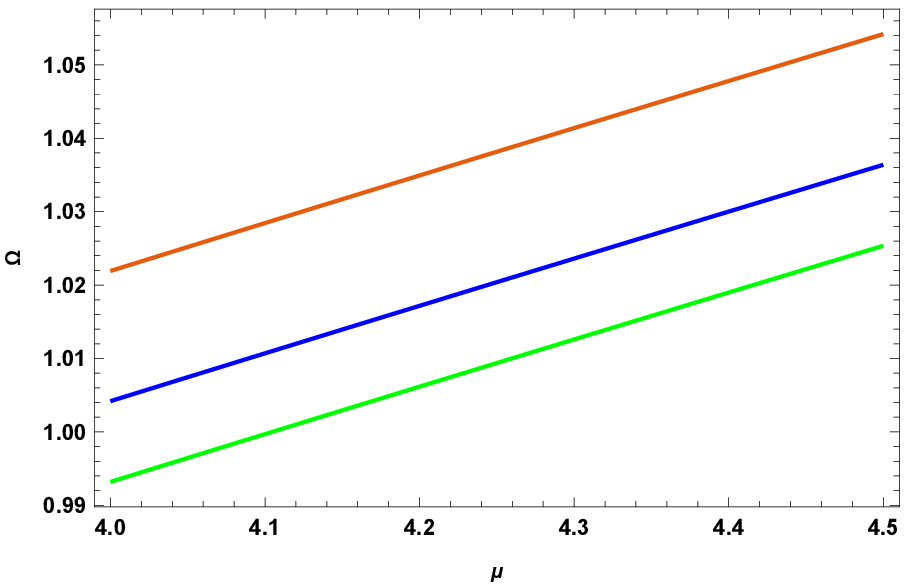}
\caption{$\Omega$ vs $\mu$ for $z=1.1$}
\end{figure}
\vspace*{-9mm}

\begin{figure}[h]
\includegraphics[scale=0.75]{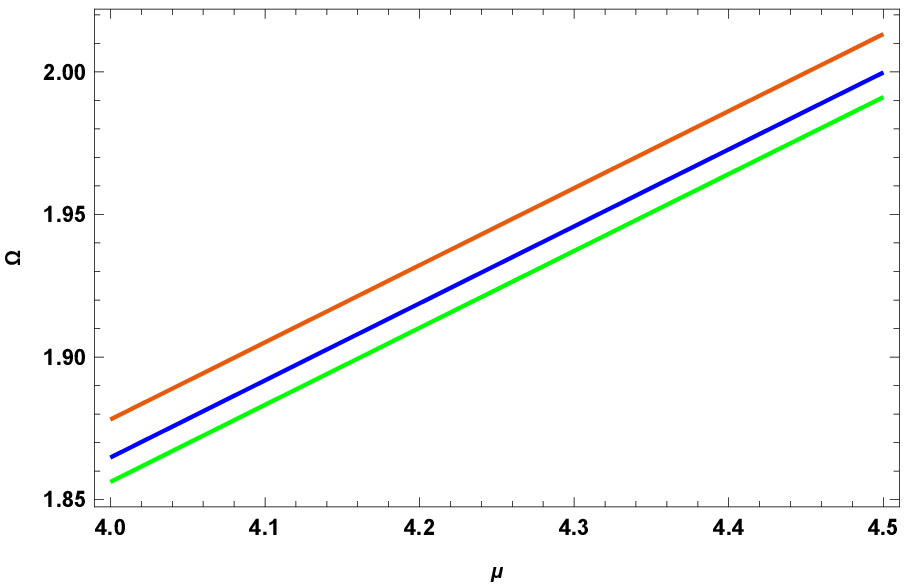}
\caption{$\Omega$ vs $\mu$ for $z=\dfrac{3}{2}$}
\end{figure}



\end{document}